\documentclass[dvips]{acta}
\usepackage{supertabular,lscape,epsfig}
\usepackage{amssymb}
\usepackage{amsmath}
\DeclareSymbolFont{ppa}{OT1}{ppl}{m}{it}
\DeclareMathSymbol{\vv}{\mathalpha}{ppa}{'166}

\newfont{\hb}{rphvb at 10pt}
\newfont{\hbo}{rphvbo at 10pt}
\newfont{\bitt}{rptmbi at 12pt}
\newfont{\bits}{rptmbi at 11pt}

\SetPages{165}{178}

\SetVol{60}{2010}

\begin{document}

\newcommand{\TabCapp}[2]{\begin{center}\parbox[t]{#1}{\centerline{
  \small {\spaceskip 2pt plus 1pt minus 1pt T a b l e}
  \refstepcounter{table}\thetable}
  \vskip2mm
  \centerline{\footnotesize #2}}
  \vskip3mm
\end{center}}

\newcommand{\TTabCap}[3]{\begin{center}\parbox[t]{#1}{\centerline{
  \small {\spaceskip 2pt plus 1pt minus 1pt T a b l e}
  \refstepcounter{table}\thetable}
  \vskip2mm
  \centerline{\footnotesize #2}
  \centerline{\footnotesize #3}}
  \vskip1mm
\end{center}}

\newcommand{\MakeTableSepp}[4]{\begin{table}[p]\TabCapp{#2}{#3}
  \begin{center} \TableFont \begin{tabular}{#1} #4 
  \end{tabular}\end{center}\end{table}}

\newcommand{\MakeTableee}[4]{\begin{table}[htb]\TabCapp{#2}{#3}
  \begin{center} \TableFont \begin{tabular}{#1} #4
  \end{tabular}\end{center}\end{table}}

\newcommand{\MakeTablee}[5]{\begin{table}[htb]\TTabCap{#2}{#3}{#4}
  \begin{center} \TableFont \begin{tabular}{#1} #5 
  \end{tabular}\end{center}\end{table}}

\newfont{\bb}{ptmbi8t at 12pt}
\newfont{\bbb}{cmbxti10}
\newfont{\bbbb}{cmbxti10 at 9pt}
\newcommand{\uprule}{\rule{0pt}{2.5ex}}
\newcommand{\douprule}{\rule[-2ex]{0pt}{4.5ex}}
\newcommand{\dorule}{\rule[-2ex]{0pt}{2ex}}
\def\thefootnote{\fnsymbol{footnote}}
\begin{Titlepage}
\Title{The Optical Gravitational Lensing Experiment.\\
The OGLE-III Catalog of Variable Stars.\\
IX.~~RR~Lyr Stars in the Small Magellanic Cloud\footnote{Based on
observations obtained with the 1.3-m Warsaw telescope at the Las Campanas
Observatory of the Carnegie Institution of Washington.}}
\Author{I.~~S~o~s~z~y~ñ~s~k~i$^1$,~~
A.~~U~d~a~l~s~k~i$^1$,~~
M.\,K.~~S~z~y~m~a~ñ~s~k~i$^1$,\\
M.~~K~u~b~i~a~k$^1$,~~
G.~~P~i~e~t~r~z~y~ñ~s~k~i$^{1,2}$,~~
\L.~~W~y~r~z~y~k~o~w~s~k~i$^3$,\\
K.~~U~l~a~c~z~y~k$^1$~~
and~~ R.~~P~o~l~e~s~k~i$^1$}
{$^1$Warsaw University Observatory, Al.~Ujazdowskie~4, 00-478~Warszawa, Poland\\
e-mail:
(soszynsk,udalski,msz,mk,pietrzyn,kulaczyk,rpoleski)@astrouw.edu.pl\\
$^2$ Universidad de Concepción, Departamento de Astronomia, Casilla 160--C, Concepción, Chile\\
$^3$ Institute of Astronomy, University of Cambridge, Madingley Road, Cambridge CB3 0HA, UK\\
e-mail: wyrzykow@ast.cam.ac.uk}
\Received{September 7, 2010}
\end{Titlepage}
\Abstract{The ninth part of the OGLE-III Catalog of Variable Stars
(OIII-CVS) comprises RR~Lyr stars in the Small Magellanic Cloud (SMC). Our
sample consists of 2475 variable stars, of which 1933 pulsate in the
fundamental mode (RRab), 175 are the first overtone pulsators (RRc), 258
oscillate simultaneously in both modes (RRd) and 109 stars are suspected
second-overtone pulsators (RRe). 30 objects are Galactic RR~Lyr stars seen
in the foreground of the SMC.

We discuss some statistical features of the sample. Period distributions
show distinct differences between SMC and LMC populations of RR~Lyr
variable stars, with the SMC stars having on average longer periods. The
mean periods for RRab, RRc and RRe stars are 0.596, 0.366 and 0.293~days,
respectively. The mean apparent magnitudes of RRab stars are equal to
19.70~mag in the {\it V}-band and 19.12~mag in the {\it I}-band. Spatial
distribution of RR~Lyr stars shows that the halo of the SMC is roughly
round in the sky, however the density map reveals two maxima near the
center of the SMC.

For each object the multi-epoch {\it V}- and {\it I}-band photometry
collected over 8 or 13 years of observations and finding charts are
available to the astronomical community from the OGLE Internet archive.}
{Stars: variables: RR~Lyrae -- Stars: oscillations -- Stars: Population II --
Magellanic Clouds}

\Section{Introduction}
RR~Lyr variable stars are low-mass radially pulsating stars with periods in
the range of 0.2--1~day. They appear in the galactic halos, thick disks and
globular clusters. RR~Lyr stars populate a narrow region in the H-R
diagram, where the horizontal branch intersects the pulsational instability
strip. Relatively small range of absolute mean magnitudes makes them useful
distance indicators. RR~Lyr stars are also excellent tracers of the oldest
observable population of stars. In account on the pulsation modes, RR~Lyr
stars can be divided into fundamental-mode (RRab), first-overtone (RRc) and
double-mode (RRd) pulsators. An existence of the second-overtone pulsators
among RR~Lyr stars (RRe) is a matter of controversy (Bono \etal 1997).

The first attempt to detect RR~Lyr stars in the Small Magellanic Cloud
(SMC) was made by Shapley (1922). He suggested that 13 variable stars in
this galaxy with periods below 1~day were cluster-type variable stars (the
historical name of RR~Lyr stars). The mean magnitude of these stars was
16.1, which was in excellent agreement with the Shapley's calibration of
the Cepheid and RR~Lyr distance scales. Years later, Payne-Gaposchkin and
Gaposchkin (1966) showed that the periods provided by Shapley (1922) for
these 13 stars were spurious, and most of these objects were classical
Cepheids. Also three short-period variable stars discovered in the SMC by
Dertayed and Landi Dessy (1952), and suggested to be RR~Lyr stars, appeared
to be classical Cepheids.

The discovery of the first actual RR~Lyr stars in the SMC was reported by
Thackeray (1951). Three RR~Lyr variable stars in the SMC cluster NGC~121
and four more in the vicinity of that cluster turned out to be of about
19~mag, which was the first independent confirmation of the revised
distance scale proposed by Baade (1952). Then, the extensive survey for
RR~Lyr stars in the SMC was performed by Graham (1975), who discovered 76
RR~Lyr stars in the field centered on NGC~121. Smith \etal (1992) found 22
certain and 20 suspected (with no period determination) RR~Lyr stars in the
Northeast Arm of the SMC. Kaluzny (1998) discovered 12 SMC RR~Lyr variable
stars behind the Galactic globular cluster 47~Tuc. At the end of the
twentieth century more than about one hundred RR~Lyr stars in the SMC were
known.

This number was significantly increased by the Optical Gravitational
Lensing Experiment (OGLE). Soszyñski \etal (2002) published the catalog of
571 RR~Lyr stars in the central 2.4 square degrees of the SMC. These
objects were discovered using the photometric material collected during the
second stage of the OGLE survey (OGLE-II). In this paper we use the
OGLE-III data to increase by several times the number of known RR~Lyr stars
in the SMC. Our catalog contains 2475 variable stars, including 30 Galactic
RR~Lyr stars in the foreground of the SMC. This is a part of the OGLE-III
Catalog of Variable Stars (OIII-CVS) which assembled so far about 130\,000
variable stars in both Magellanic Clouds. Among others, we published the
catalog of 24\,906 RR~Lyr stars in the Large Magellanic Cloud (LMC;
Soszyñski \etal 2009), and the catalogs of classical (4630 objects) and
type~II (43 objects) Cepheids in the SMC (Soszyñski \etal 2010ab).

The paper is structured as follows. In Section~2 we present the photometric
data and the reduction methods. Section~3 describes the variable stars
selection process. Section~4 presents the catalog itself. Section~5 is
devoted to the comparison of our sample with the previously published
catalogs of RR~Lyr stars in the SMC. Finally, we discuss some features of
our sample of RR~Lyr stars in Section~6.

\vspace*{9pt}
\Section{Observational Data}
\vspace*{3pt}
The catalog is based on the observations carried out with the 1.3-m Warsaw
telescope at the Las Campanas Observatory, Chile. The telescope was
equipped with the eight-chip CCD mosaic camera of the total resolution
$8192\times8192$ pixels and the field of view of about
$35\times35.5$~arcmin. For details of the instrumental setup we refer to
Udalski (2003).

The OGLE-III project obtained time-series observations of the SMC during
about 800 nights between June 2001 and May 2009. About 90\% of the
observations were made with the {\it I}-band filter, the remaining
measurements were obtained in the {\it V} photometric band. The exposure
time was 180~sec and 225~sec in the {\it I}- and {\it V}-bands,
respectively. The total area regularly observed by the OGLE-III project was
about 14 square degrees distributed over 41 fields. The OGLE data reduction
pipeline was developed by Udalski (2003) on the basis of the Difference
Image Analysis (DIA; Alard and Lupton 1998, Alard 2000, Wo¼niak 2000). Full
description of the reduction techniques, photometric calibration and
astrometric transformations can be found in Udalski \etal (2008a).

Some stars were detected twice, because they were located in the
overlapping regions of adjacent fields. The photometry of these objects was
compiled from all available sources. For the stars in the central 2.4
square degrees of the SMC the OGLE-III photometry was supplemented with the
OGLE-II observations (Szymañski 2005) collected between 1997 and 2000. Both
datasets were tied by shifting the OGLE-II photometry to agree with the
OGLE-III light curves.

\vspace*{9pt}
\Section{Selection and Classification of RR~Lyr Stars}
\vspace*{3pt}
All {\it I}-band light curves collected by the OGLE-III survey were
searched for periodicity with the {\sc Fnpeaks} code kindly provided by
Z.\, Ko³aczkowski. The software used Discrete Fourier Transform to output
the most significant periodicities with amplitudes and signal-to-noise
ratios ($S/N$) for a given light curve. The Fourier spectra were calculated
up to 24~d$^{-1}$ for each star. All light curves with $S/N>5$ and primary
periods between 0.2 and 1~day were visually examined and divided into
pulsating-like, eclipsing-like and other variable objects. Then, the stars
tentatively classified as possible pulsating variable stars were classified
by means of the detected periods, colors, magnitudes, amplitudes, light
curve shapes, period ratios (for multiperiodic stars), etc.

\begin{figure}[htb]
\hglue-10mm{\includegraphics[width=14.2cm, bb=10 265 585 745]{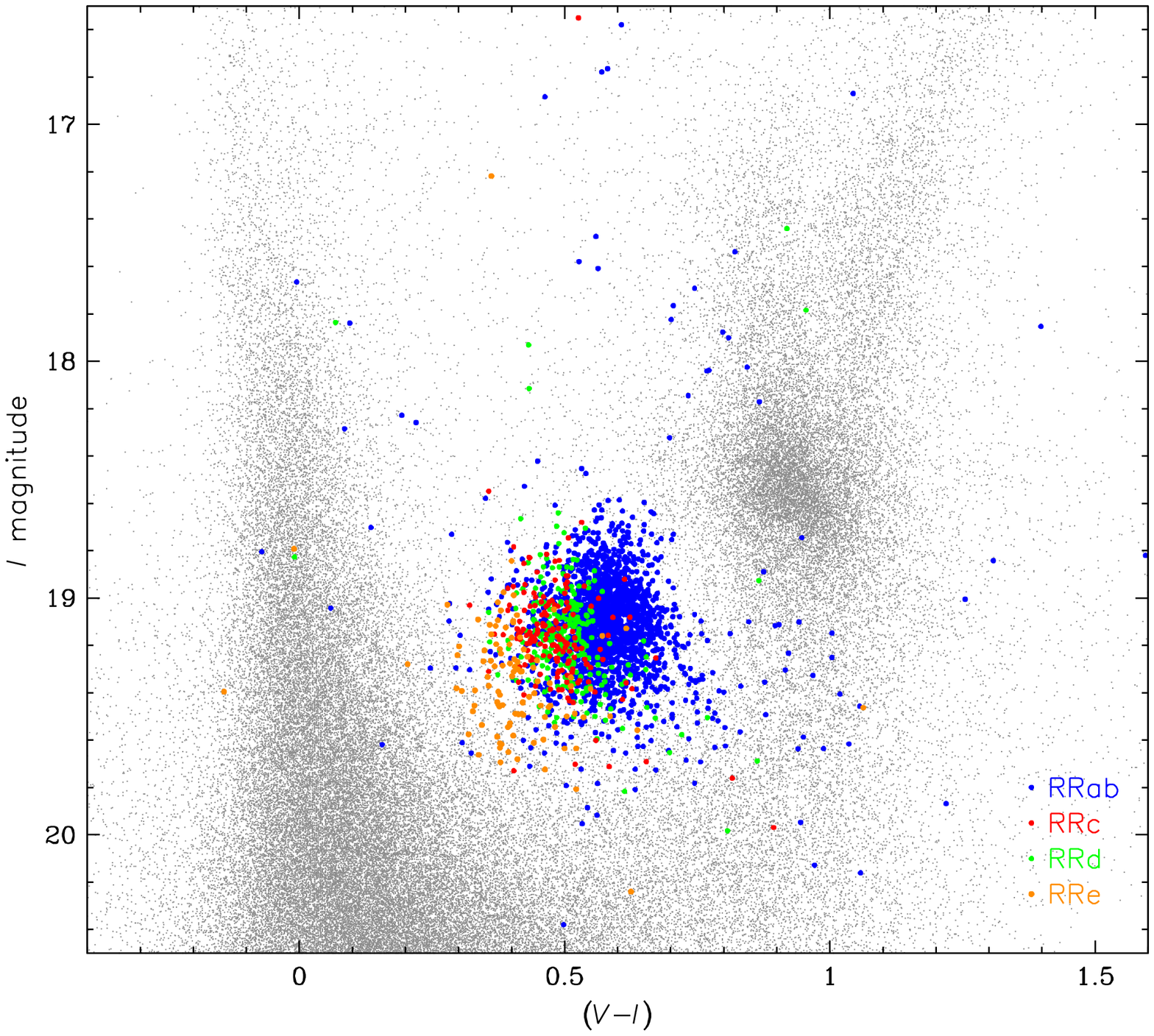}}
\FigCap{Color--magnitude diagram for RR~Lyr stars in the SMC. Blue points
represent RRab stars, red -- RRc stars, green -- RRd stars, yellow -- RRe
stars. Grey dots show all stars from the fields SMC100.1 and SMC100.2.}
\end{figure}

Our search revealed a number of $\delta$~Sct stars, which lie at the
extension of the period--luminosity relation for first-overtone classical
Cepheids, so they are generally brighter than RR~Lyr stars with the same
periods. The list of $\delta$~Sct stars in the SMC will be published
elsewhere. Some of the stars categorized in our catalog as RR~Lyr stars
have atypical $(V-I)$ colors and magnitudes in one or both filters. These
objects are visible as outliers in the $(V-I)$ \vs $I$ color--magnitude
diagram plotted in Fig.~1. The amplitudes of these variable stars are
usually reduced compared to typical RR~Lyr stars with the same periods. We
assumed that these objects are unresolved blends of RR~Lyr variable stars
with other stars. In these cases our classification is based on their light
curve morphology. Our sample includes also 30 variable stars brighter than
SMC RR~Lyr stars that are likely Galactic RR~Lyr stars in the foreground of
the SMC. For several of the brightest (\ie the nearest) objects from that
group proper motions are detectable in the OGLE data. One of the foreground
RR~Lyr stars (OGLE-SMC-RRLYR-0051 = V9 = HV~810) belongs to 47~Tuc cluster
(Storm \etal 1994).

\begin{figure}[htb]
\hglue-6mm{\includegraphics[width=13.7cm, bb=10 400 585 745]{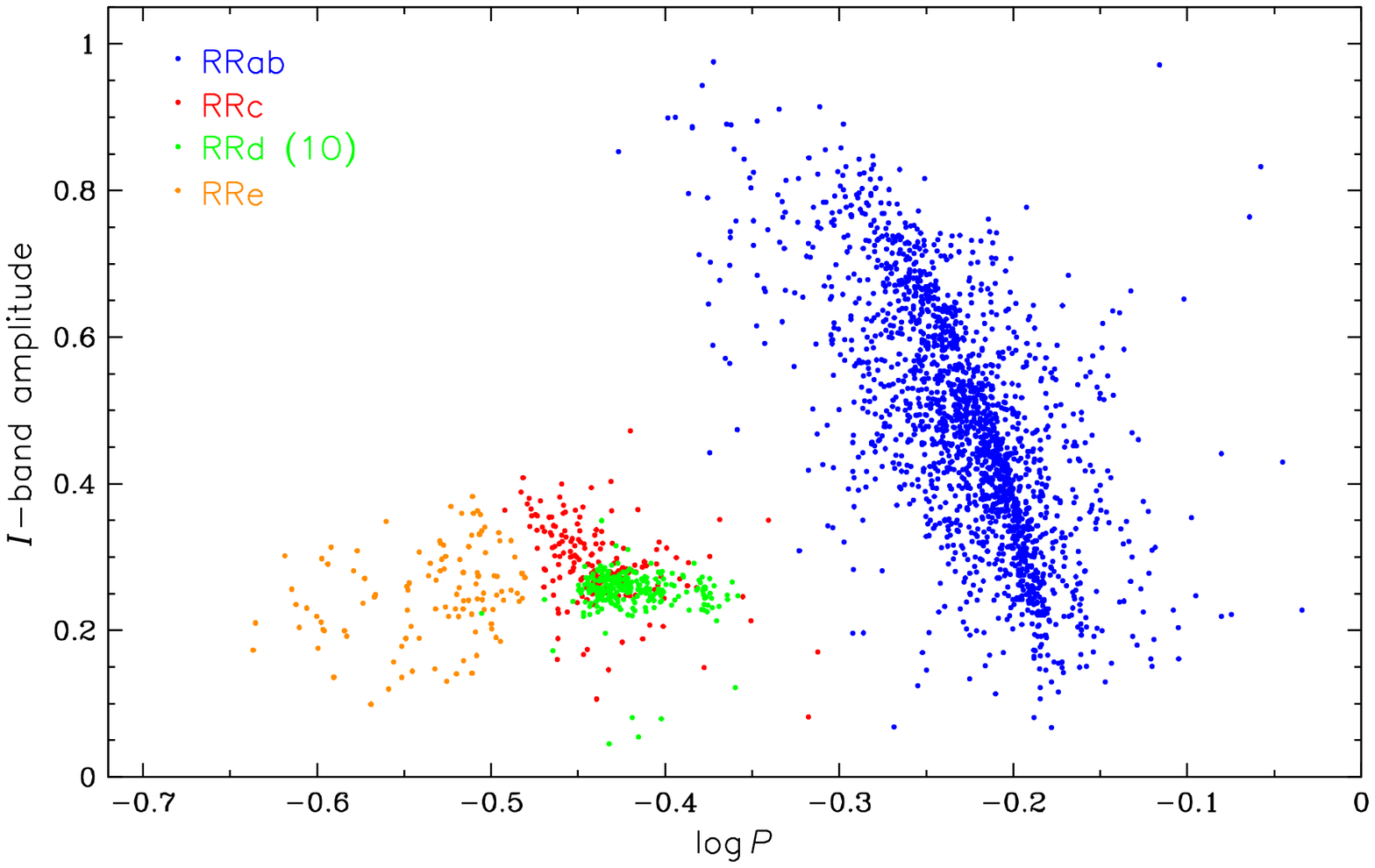}}
\FigCap{Period--amplitude diagram for RR~Lyr stars in the SMC. Color
symbols denote the same type of stars as in Fig.~1. RRd stars are
represented by the first-overtone periods.}
\end{figure}

Single-mode RR~Lyr stars were divided into three groups: RRab, RRc and RRe
variable stars. Distinguishing between fundamental-mode and overtone
pulsators was a relatively easy task, thanks to the very different shapes
of the light curves in both classes. In several questionable cases we also
used the amplitudes of variations (Fig.~2), positions in the
period--luminosity diagrams and parameters of the Fourier light curve
decomposition. We decided to separate RRc and RRe variable stars, because,
like in the LMC (Soszyñski \etal 2009), we noticed additional peak in the
period distribution for the shortest-period RR~Lyr stars (Fig.~3). The peak
is centered at $P\approx0.31$~days. Similar secondary maximum was observed
also in the Sculptor dwarf spheroidal galaxy (Kaluzny \etal 1995). This
peak is usually attributed to the second-overtone RR~Lyr stars (\eg Alcock
\etal 1996), although other explanations are also possible (\eg Bono \etal
1997 suggested that this peak is a signature of a metal-rich population of
RR~Lyr stars). We separated RRe stars from RRc variable stars using their
positions in the period--amplitude diagram (Fig.~2). The limiting maximum
period of RRe stars was adopted at $P=0.33$~days. Such an approach
certainly has only statistical meaning. In individual cases our distinction
between RRc and RRe stars may be wrong.

\begin{figure}[htb]
\hglue-10mm{\includegraphics[width=14.2cm, bb=0 175 585 755]{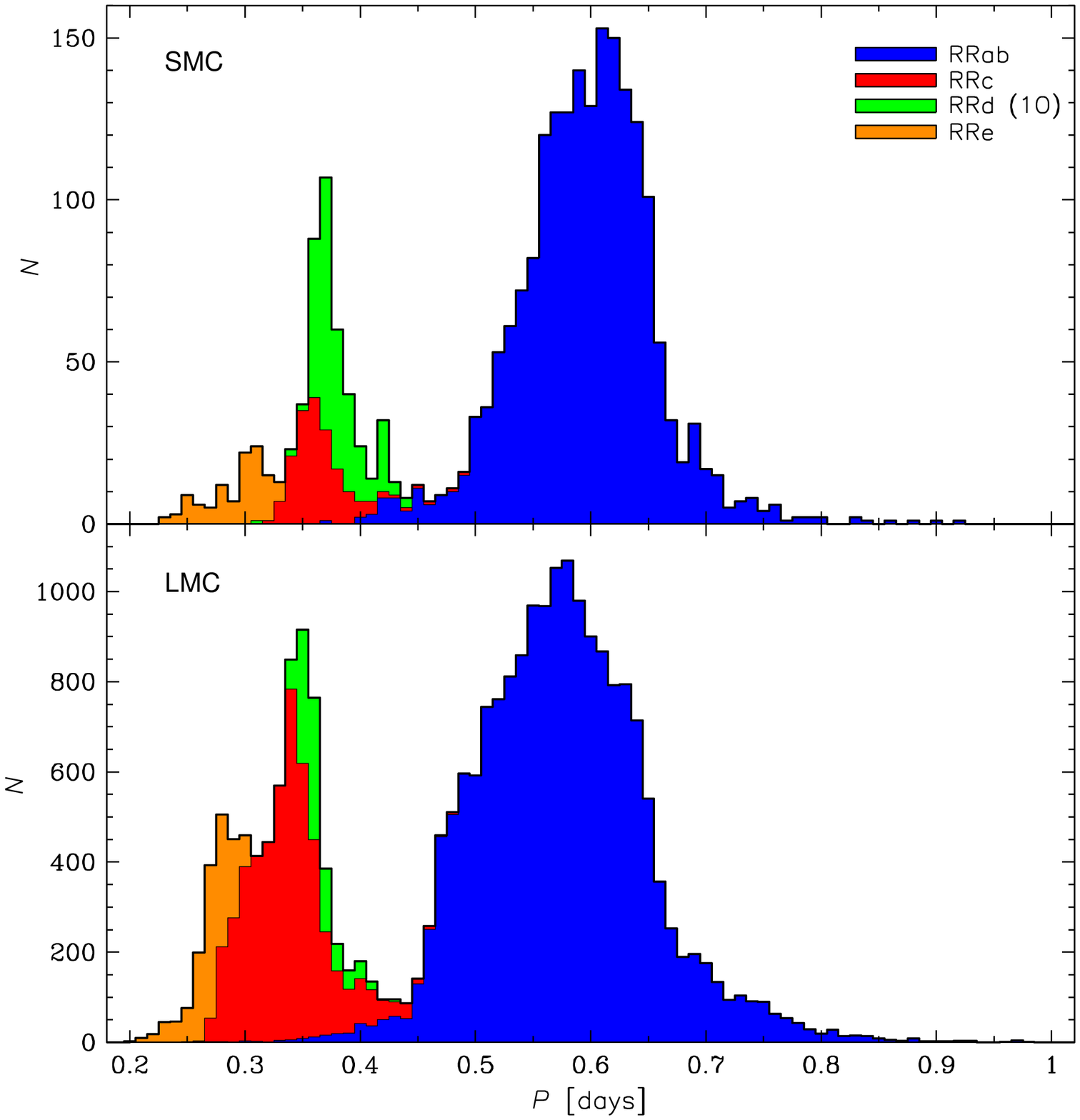}}
\FigCap{Period distribution for RR~Lyr stars in the SMC and LMC: cumulative
histogram with different types of RR~Lyr stars shown with different
colors. Blue regions show RRab stars, red -- RRc stars, green -- the
first-overtone period of RRd stars, and orange -- RRe stars. The width of
bins is 0.01~day.}
\end{figure}

\begin{figure}[htb]
\centerline{\includegraphics[width=15.5cm, bb=20 375 575 745]{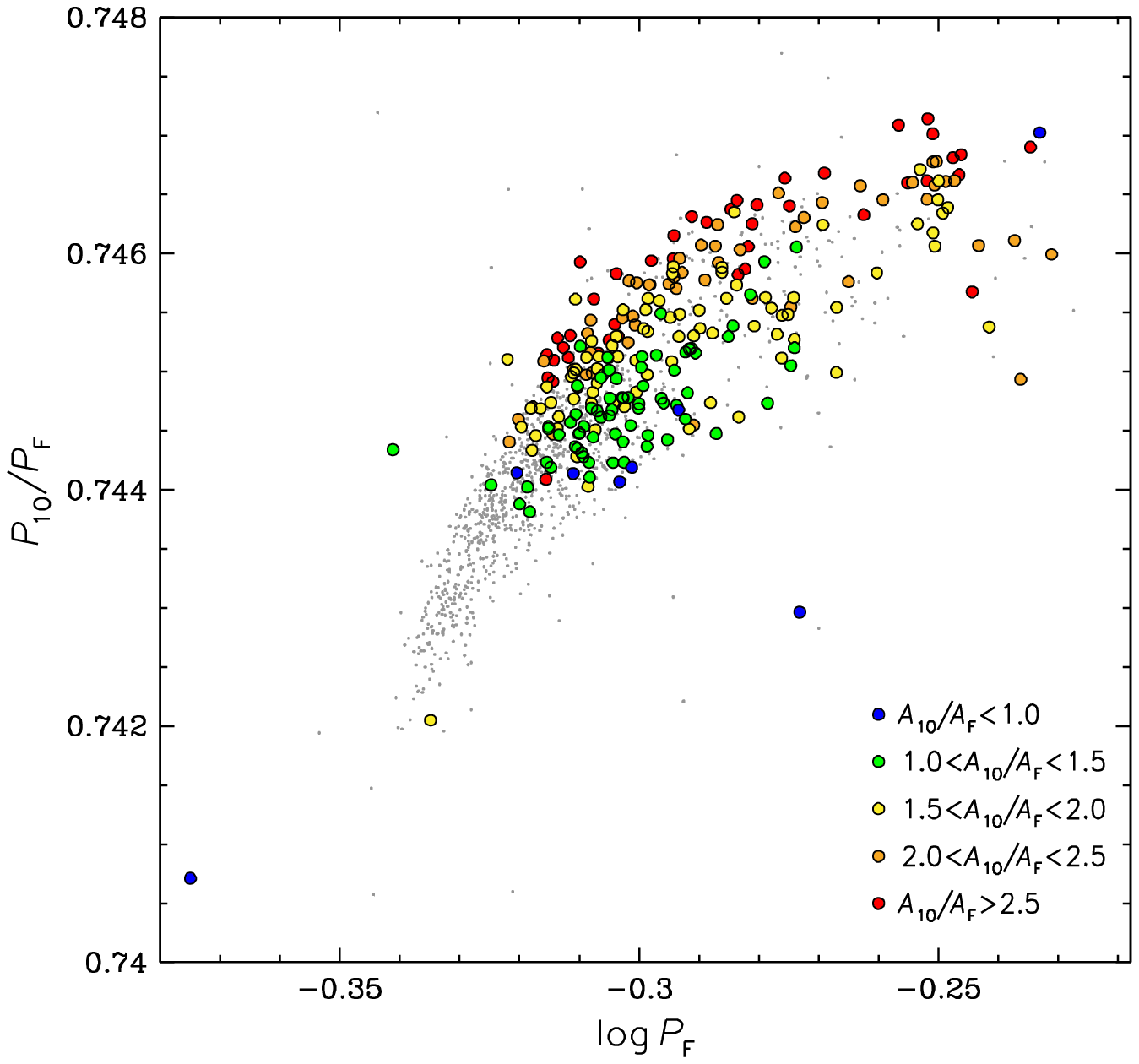}}
\vspace*{-7pt}
\FigCap{Petersen diagram for RRd stars in the SMC. Different colors
represent different amplitude ratios between the first overtone and
fundamental mode. Grey small dots show RRd stars in the LMC (Soszyñski
\etal 2009).}
\end{figure}

We performed a search for double-mode RR~Lyr stars (RRd) in two
ways. First, we carried out a search for the secondary periods for all
previously identified RR~Lyr stars. Each light curve was fitted with a
Fourier series with a number of harmonics minimizing the $\chi^2$ per
degree of freedom, the function was subtracted from the observational data
and the period search was performed on the residual data. Then, the stars
with periods and period ratios characteristic for RRd variable stars (\ie
with longer periods in the range 0.42--0.6~days and shorter-to-longer
period ratios between 0.74 and 0.75) were subjected for visual
inspection. From that group we selected stars with significant secondary
periods ($S/N>4$).

The second method used to select double-mode RR~Lyr stars based on the
massive period search performed for all stars observed in the SMC by the
OGLE-III project. We checked two dominant periods obtained from this
analysis, searching for objects with periods and period ratios typical for
RRd stars. After careful visual inspection of the light curves we extended
the previously selected sample of RRd stars by a few objects. In total, we
found 258 RRd stars. Fig.~4 shows the Petersen diagram (\ie a plot of the
period ratio \vs logarithm of the fundamental period) for our sample. For
comparison we also present here the LMC RRd stars from the catalog of
Soszyñski \etal (2009). Different colors of the points show different
amplitude ratio of both pulsation modes ($A_{1O}/A_{\rm F}$). It is clear
that the ratio of amplitudes is strongly correlated with the position of a
double-mode RR~Lyr star in the Petersen diagram.

During the search for RRd stars we noticed objects with the secondary
periods very similar to the primary ones. Such a phenomenon can be related
to the Blazhko effect or to the changes of the primary period. The time
baseline of the OGLE data span up to 13 years, so the period changes are
detectable for some RR~Lyr stars. We detected closely-spaced secondary
frequencies for about 22\% of RRab stars, and 14\% of the (single-mode)
overtone variables.

\begin{figure}[htb]
\hglue-1mm{\includegraphics[width=13.1cm]{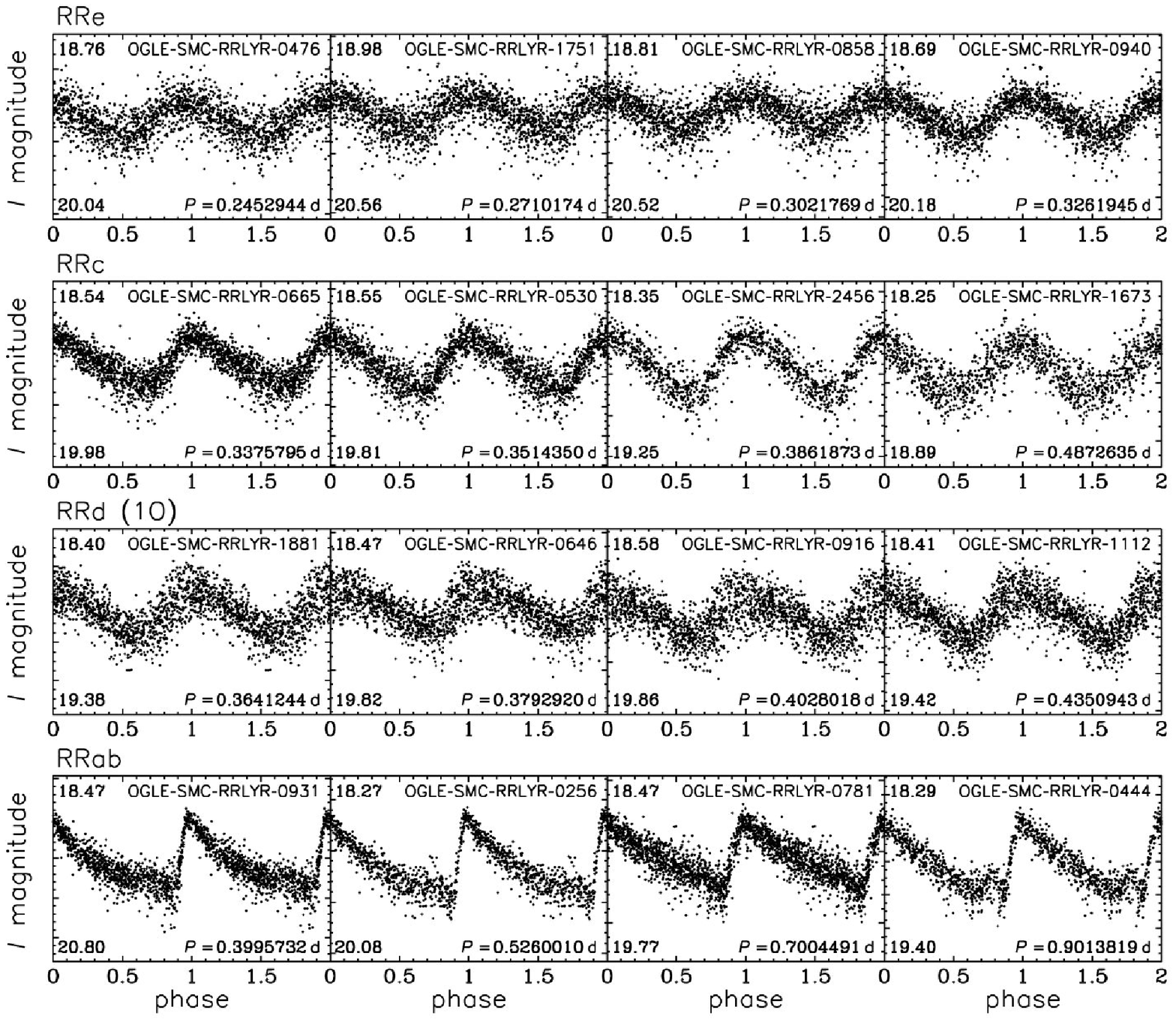}}
\vspace{1mm}
\FigCap{Representative light curves of RR~Lyr stars from our catalog. First
row presents four RRe variable stars, second row -- RRc stars, third row --
RRd stars phased with the first-overtone period, and the last row shows
light curves of RRab star. The range of magnitudes varies from panel to
panel. Numbers in the left corners show the lower and the upper limits of
magnitudes.}
\end{figure}

Representative light curves of RR~Lyr variable stars of various types are
presented in Fig.~5. The scatter of the light curves caused by the
photometric errors is substantial, as RR~Lyr stars in the SMC are close to
the detection limit of the OGLE-III SMC photometry. However, generally
there are no problems with the variability type identification. Some stars
with uncertain classification are flagged in the remarks of the catalog.

\Section{Catalog of RR~Lyr Stars in the SMC}
We identified in total 2475 RR~Lyr stars in the SMC OGLE-III fields. The
sample consists of 1933 RRab, 175 RRc, 258 RRd and 109 RRe stars. 30
objects are Galactic foreground RR~Lyr stars, of which 22 pulsate in the
fundamental mode, one was classified as an RRc star, three as RRe stars and
four are double-mode RR~Lyr variable stars. The catalog data are available
through the WWW interface or from the anonymous FTP site:
\begin{center}
{\it http://ogle.astrouw.edu.pl/} \\ {\it
ftp://ftp.astrouw.edu.pl/ogle/ogle3/OIII-CVS/smc/rrlyr/}\\
\end{center}

The stars arranged in order of right ascension are listed in the file {\sf
ident.dat} at the FTP site. The object designation (in the form
OGLE-SMC-RRLYR-NNNN, where NNNN is a four-digit consecutive number),
OGLE-III field and internal database number (consistent with the
photometric maps of the SMC by Udalski \etal 2008b), mode of pulsation
(RRab, RRc, RRd, RRe), equinox J2000.0 right ascension and declination,
cross-identifications with the OGLE-II photometric database (Szymañski
2005), cross-identifications with the extragalactic part of the General
Catalogue of Variable Stars (GCVS; Artyukhina \etal 1995), and other
designations are given in the file {\sf ident.dat}.

Observational parameters of the stars -- intensity-averaged $\langle
I\rangle$ and $\langle V\rangle$ magnitudes, periods with uncertainties
(derived with the {\sc Tatry} code developed by Schwarzenberg-Czerny 1996),
peak-to-peak {\it I}-band amplitudes and parameters of the Fourier light
curve decomposition (Simon and Lee 1981) -- are provided in the files {\sf
RRab.dat}, {\sf RRc.dat}, {\sf RRd.dat} and {\sf RRe.dat}. Additional
information on some objects can be found in the file {\sf remarks.txt}. The
OGLE-II and OGLE-III multi-epoch {\it VI} photometry can be downloaded from
the directory {\sf phot/}. Finding charts for each star are stored in the
directory {\sf fcharts/}. These are $60\arcs\times60\arcs$ subframes of the
{\it I}-band DIA reference images.

\Section{Cross-Identification with Other Catalogs}
To test the completeness of our catalog we cross-matched our sample with
the previously released lists of RR~Lyr variable stars in the SMC. The
largest catalog of these objects published heretofore was the OGLE-II
catalog (Soszyñski \etal 2002). We independently identified 558 stars from
571 objects classified as RR~Lyr variable stars in the OGLE-II catalog of
RR~Lyr stars in the SMC. We checked the missed 13 objects and noticed that
eight of them have been reclassified as $\delta$~Sct stars, short-period
classical Cepheids or artifacts. The remaining five stars turned out to be
RR~Lyr stars, usually of RRc type. They were not classified as RR~Lyr stars
in the initial classification procedure due to a small number of observing
points in the OGLE-III database or atypical colors. We included all the
missing RR~Lyr stars in the present catalog.

\begin{figure}[p]
\centerline{\includegraphics[width=15.0cm, bb=0 50 560 745]{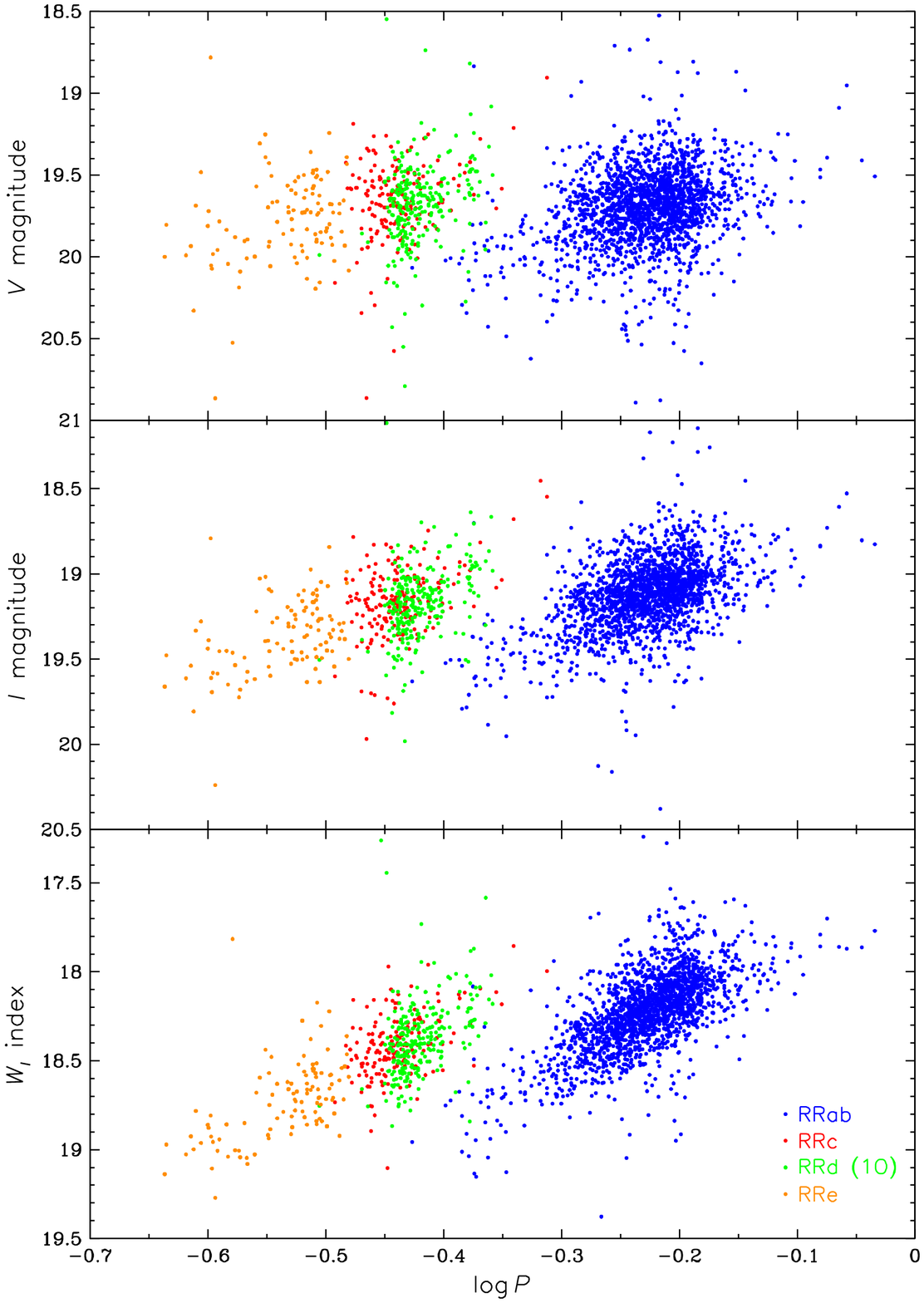}}
\FigCap{Period--luminosity relations for RR~Lyr in the SMC. Color symbols
represent the same type of stars as in Fig.~1. {\it Upper panel} presents
$\log{P}$ \vs $V$ diagram, {\it middle panel} -- $\log{P}$ \vs $I$ diagram,
{\it lower panel} -- $\log{P}$ \vs $W_I$ diagram, where $W_I=I-1.55(V-I)$ is
the extinction insensitive Wesenheit index.}
\end{figure}

Most of the RR~Lyr stars in the SMC provided by the GCVS Vol.~V (Artyukhina
\etal 1995) originate from the paper by Graham (1975). Unfortunately, the
majority of these objects lie outside the OGLE-III fields. From 27 RR~Lyr
stars that may be potentially found in the region monitored by the OGLE-III
survey, we identified practically all objects. Only one star -- BV~Tuc --
was reclassified as a classical Cepheid. Our catalog includes also six
Galactic RR~Lyr stars that were incorrectly classified in the GCVS as
BL~Boo type stars (\ie anomalous Cepheids). From the list of RR~Lyr stars
in the vicinity of 47~Tuc published by Weldrake \etal (2004), we identified
19 objects, \ie all stars covered by the OGLE-III fields.

\Section{Discussion}
A large number of RR~Lyr stars in the SMC allows us to perform a
preliminary analysis of their statistical properties, and to compare them
to a huge sample of RR~Lyr variable stars in the LMC identified by the OGLE
project (Soszyñski \etal 2009). Period--amplitude diagram for the SMC
sample is shown in Fig.~2. Comparing to the LMC and the field of the
Galaxy, RR~Lyr stars in the SMC seem to constitute much more homogeneous
group in terms of metallicity, history of star formation and reddening. In
particular, Galactic field RR~Lyr stars manifest clear Oosterhoff dichotomy
(Suntzeff \etal 1991, Szczygie³ \etal 2009), which is not detectable in the
SMC.

Fig.~3 presents the period distribution of the SMC and LMC RR~Lyr stars.
These are cumulative histograms with different types of RR~Lyr stars
plotted in different colors. It is striking that both galaxies host
different proportion of variable stars pulsating in various modes. In the
LMC, RRc and RRe stars constitute together 25\% of all RR~Lyr stars, but
the whole population comprises only 4\% of RRd stars. In the SMC
single-mode overtone pulsators (RRc and RRe stars) constitute only 11.5\%
of the RR~Lyr sample, and there is a similar number of RRd stars. Such a
discrepancy cannot be explained by a possible incompleteness of the SMC
catalog relative to the LMC sample.

\begin{figure}[p]
\centerline{\includegraphics[width=12.1cm]{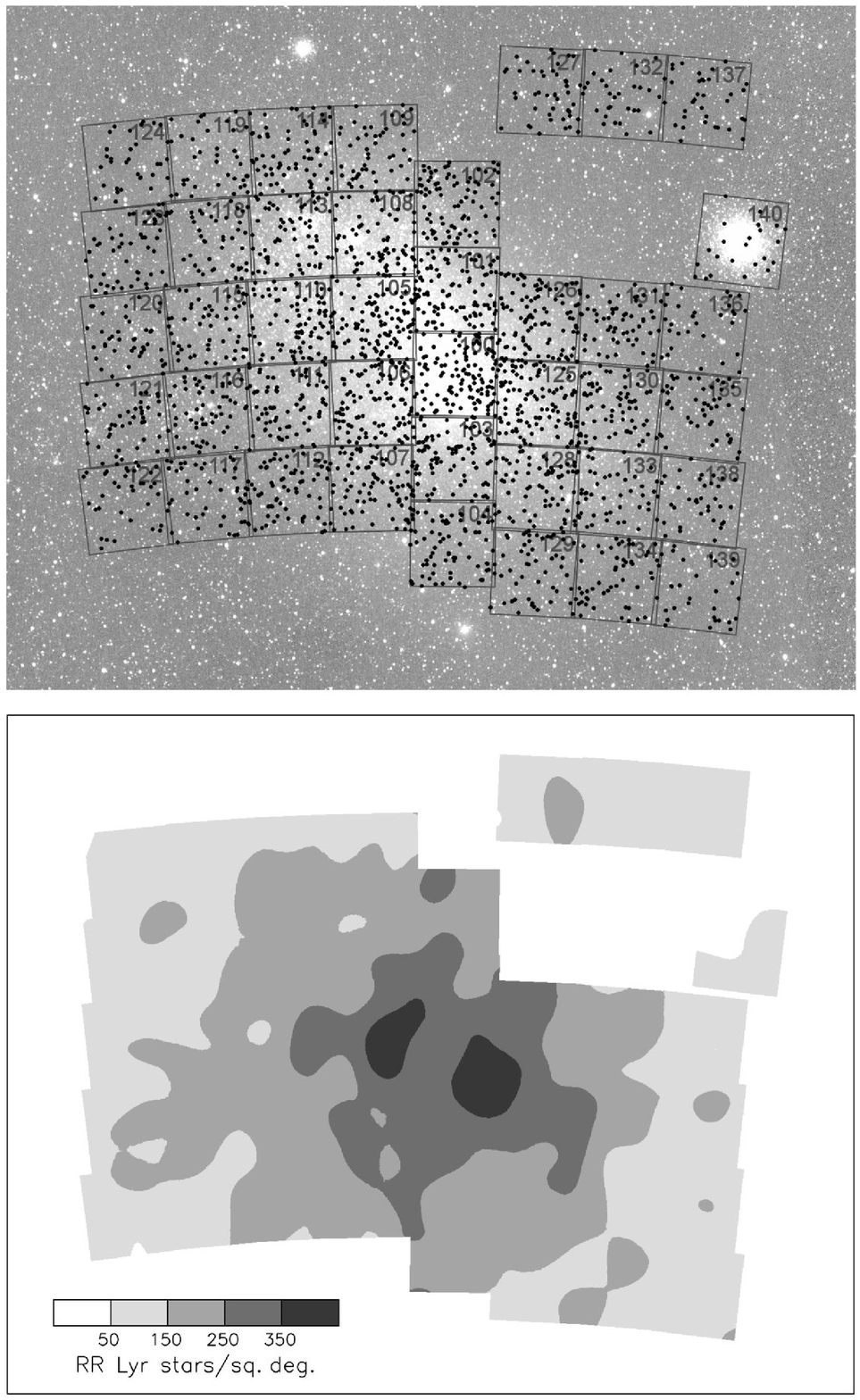}}
\vspace{2mm}
\FigCap{{\it Upper panel}: spatial distribution of RR~Lyr stars in the
SMC. The background image of the SMC originates from the ASAS sky survey
(Pojmañski 1997). {\it Lower panel}: surface density map of RR~Lyr stars in
the SMC.}
\end{figure}

\MakeTable{l@{\hspace{10pt}}
c@{\hspace{7pt}} c@{\hspace{12pt}} c@{\hspace{5pt}} c}{12.5cm}{Mean and
Modal Periods of RR~Lyr Stars in the SMC and LMC.}  {\hline
\noalign{\vskip3pt}
 & \multicolumn{2}{c}{SMC~~~~~~~} & \multicolumn{2}{c}{LMC~~~} \\ & mean
 period & modal period & mean period & modal period \\ & [days] & [days] &
 [days] & [days] \\
\noalign{\vskip3pt}
\hline
\noalign{\vskip3pt}
RRab     & 0.596 & 0.615 & 0.576 & 0.580 \\ RRc      & 0.366 & 0.366 &
0.337 & 0.341 \\ RRd (1O) & 0.380 & 0.369 & 0.363 & 0.357 \\ RRe      &
0.293 & 0.309 & 0.270 & 0.272 \\
\noalign{\vskip3pt}
\hline}

Fig.~3 shows that RR~Lyr stars in the Large and Small Magellanic Clouds
reveal different mean and modal periods. Table~1 compares the typical
periods for various types of RR~Lyr variable stars in both
environments. The SMC variable stars have average periods by
0.02--0.03~days longer than the LMC RR~Lyr stars. Mean apparent unreddened
magnitudes of RRab stars (after removing Galactic and blended objects) are
equal to 19.12~mag in the {\it I}-band and 19.70~mag in the {\it V}-band,
however a weak period--luminosity relation is visible in both filters
(Fig.~6).

In Fig.~7 we present the spatial distribution of RR~Lyr stars in the
SMC. Upper panel shows the position of individual objects overplotted on
the image of the SMC obtained by the ASAS project (Pojmañski 1997). Lower
panel displays the density map of the SMC RR~Lyr stars derived by smoothing
the above distribution with the Gaussian filter. It is clear that the
population of RR~Lyr stars in the SMC is distributed over large area, much
larger than 14~square degrees covered by the OGLE-III fields. We expect
that the current phase of the OGLE project (OGLE-IV) will cover virtually
all RR~Lyr stars in both Magellanic Clouds.

RR~Lyr in the SMC form roughly a round structure in the sky, in contrast to
the LMC (Soszyñski \etal 2009), where the spatial distribution of RR~Lyr
stars is not spherical and is elongated in the same direction as the LMC
bar. Pejcha and Stanek (2009) used our catalog to study 3D structure of the
LMC halo. They noticed that the RR~Lyr distribution can be approximated by
a triaxial ellipsoid with the longest axis almost parallel to the line of
sight. The scatter of points in the reddening-free period--Wesenheit index
diagram (Fig.~6) is considerably larger for RR~Lyr stars in the SMC than in
the LMC, so the SMC stellar halo is probably more extended along the line
of sight than in the LMC. The standard deviation of residuals after
subtracting the least-square fit to the $\log{P}{-}W_I$ relation is equal
to 0.11~mag in the LMC, while in the SMC it exceeds 0.14~mag. It is
interesting that the spatial distribution of RR~Lyr stars in the SMC
(Fig.~7) seems to have two maxima, around the coordinates
$\alpha_{\rm J2000}=00\uph47\zdot\upm4$, $\beta_{\rm J2000}=-73\arcd12\arcm$ and
$\alpha_{\rm J2000}=00\uph56\zdot\upm7$, $\beta_{\rm J2000}=-72\arcd54\arcm$.

\Acknow{We thank W.~Dziembowski for valuable comments and suggestions
which helped to improve the paper. We are grateful to Z.~Ko³aczkowski,
G.~Pojmañski, A.~Schwarzenberg-Czerny and J.~Skowron for providing codes
and data which enabled us to prepare this study.

This work has been supported by MNiSW grant N~N203~2935~33. The research
leading to these results has received funding from the European Research
Council under the European Community's Seventh Framework Programme
(FP7/2007-2013)/ERC grant agreement no. 246678. The massive period search
was performed at the Interdisciplinary Centre for Mathematical and
Computational Modeling of Warsaw University (ICM), project no.~G32-3. We
wish to thank M. Cytowski for his skilled support.}

\end{document}